\documentclass[showpacs,showkeys,preprint]{revtex4}
\usepackage{latexsym}
\usepackage{amsmath}
\usepackage{amssymb}
\usepackage{amsthm}
\usepackage{graphicx}
\usepackage{dcolumn}
\newcommand{\be}{\begin{equation}}
\newcommand{\ee}{\end{equation}}
\newcommand{\ba}{\begin{eqnarray}}
\newcommand{\ea}{\end{eqnarray}}

\begin{document}
\title{Reentrant melting of the exp-6 fluid: \\
the role of the repulsion softness}
\author{Franz Saija$^1$~\cite{aff1}, Gianpietro Malescio$^2$~\cite{aff2},
and Santi Prestipino$^1$~\cite{aff3}}
\affiliation{
$^1$ CNR-Istituto per i Processi Chimico-Fisici, Contrada Papardo,
98158 Messina, Italy \\
$^2$ Universit\`a degli Studi di Messina, Dipartimento di Fisica,
Contrada Papardo, 98166 Messina, Italy}
%\date{\today}
\begin{abstract}
We investigate the phase behaviour of a system of particles interacting
through the exp-6 pair potential, a model interaction that is appropriate
to describe effective interatomic forces under high compression.
The soft-repulsive component of the potential is being varied so as
to study the effect on reentrant melting and density anomaly.
Upon increasing the repulsion softness, we find that the anomalous
melting features persist and occur at smaller pressures.
Moreover, if we reduce the range of downward concavity in the potential
by extending the hard core at the expenses of the soft-repulsive shoulder,
the reentrant part of the melting line reduces in extent so as it does
the region of density anomaly.
\end{abstract}
\pacs{61.20.Ja, 62.50.-p, 64.70.D-}
\keywords{High-pressure phase diagrams of the elements, Liquid-solid
transitions, Reentrant melting}
\maketitle

\section{Introduction}

At high pressures, a number of elements in the periodic table show a
maximum in the fluid-solid coexistence temperature, followed by a region
of reentrant melting, see e.g. Cs, Rb, Na, Ba, Te, etc.~\cite{young1}
A further pressure increase makes the slope of the melting line positive
again. This behaviour has been called ``anomalous'' as opposed to the
``standard'' behaviour typical of simple fluids, consisting in a regularly
increasing and concave melting curve.
The class of substances exhibiting anomalous melting constantly expands
as advances in experimental methods allow to reach higher and higher
pressures. Anomalous melting has been related to a certain degree of
softness, as induced by pressurization, of the interatomic
repulsion~\cite{young2,mcmillan,ross,prestipino1}. A similar behaviour
is observed in a completely different type of systems, i.e., polymer
solutions and colloidal
dispersions~\cite{royall,likos1,likos2,malescio1}, where it usually appears
in combination with other water-like anomalies~\cite{stanley,deoliveira1}.
In order to account for these melting oddities, numerous effective pair
potentials were proposed in the past, some of them being even bounded at
zero separation~\cite{stell,young2,stillinger,prestipino2,watzlawek,quigley,gibson,deoliveira2,deoliveira3}.
Core-softened potentials generally present a region of downward concavity
in their repulsive component~\cite{debenedetti}.

A classical spherically-symmetric potential that is widely used in the
realm of high-pressure physics is the Buckingham or exp-6
potential~\cite{buckingham,ross}, where the short-range repulsion is
modelled through a hard-core plus a soft-repulsive exponential shoulder:
\be
u(r)=\left\{
\begin{array}{ll}
+\infty & ,\,\,\,r<\sigma_M \\
\frac{\epsilon}{\alpha-6}\left[
6\,e^{-\alpha\left(r/\sigma-1\right)}-
\alpha\left(\frac{\sigma}{r}\right)^6\right] & ,\,\,\,r\ge\sigma_M
\end{array}
\right.
\label{eq1}
\ee
Here $r$ is the interparticle distance, $\epsilon$ is the depth of the
attractive well, $\sigma$ is the position of the well minimum, $\alpha$
(usually taken in the range 10-15 ~\cite{fried}) controls the steepness
of the exponential repulsion, and $\sigma_M(\alpha)$ is the point where the
function in the second line of Eq.\,(\ref{eq1}) attains its maximum value.
The exp-6 potential satisfies the core-softening condition,
i.e., there exists a range of interparticle distances where the repulsive 
force decreases as two particles get closer to each other~\cite{malescio2}.
This gives origin to two separate repulsive length scales, i.e., a larger
one associated with the soft repulsion (being effective at the lower
pressures) and a smaller one related to the particle-core diameter
$\sigma_M$ (dominant at higher pressures). It has been shown that, upon
increasing the pressure $P$, the melting temperature of the exp-6 system
passes through a maximum followed by a region of
reentrant melting; upon further compression, the melting line eventually
recovers a positive slope~\cite{malescio2}. This behaviour is related to the
existence of two different patterns of short-range order in the system: 
an open one (associated with the soft-repulsive scale) and a compact one
(associated with the hard core). 
The reentrance of the fluid phase at intermediate pressures (and for
not too low temperature) follows from the packing frustration induced
by the interplay between these two local structures.

To better understand the role of the soft repulsion for the occurrence of
anomalous melting, we investigate how the phase behaviour of the exp-6 model
changes when varying the softness of the potential. This variation can
be achieved in two ways. The most direct one is by changing
the exponent $\alpha$ controlling the steepness of the exponential repulsion.
A different way of varying the degree of softness is to define, for fixed
$\alpha$, a whole sequence of modified exp-6 interactions by shifting to
higher and higher distance the point where the repulsion changes from
hard-core to exponential. For the original exp-6 interaction, this crossing
point occurs at $r_{cross}=\sigma_M$; taking $r_{cross}$ to be larger than
$\sigma_M$, the ensuing repulsion turns out to be softer than the exp-6 law.

The paper is organized as follows: In Section 2, we introduce the numerical
approach that is used to construct the phase diagram of the system.
Section 3 is devoted to a discussion of the results while further remarks
and conclusions are deferred to Section 4.

\section{Monte Carlo simulation}

We perform Monte Carlo (MC) simulations of the exp-6 model in the
isothermal-isobaric ($NPT$) ensemble, using the standard Metropolis
algorithm and periodic boundary conditions. The simulations are carried
out for a number of $N=432$ (bcc) and 500 particles (fcc) in a cubic
box (we checked that finite-size effects are negligible).
For each pressure $P$ and temperature $T$, equilibration of the sample
typically takes some $10^4$ MC sweeps, a sweep consisting of $N$ attempts
to change the position of a random particle, followed by one attempt
to modify the box volume.
The maximum random displacement of a particle and the maximum volume
update in a trial MC move are adjusted every sweep during the run so
as to keep the acceptance ratio of the moves as close to $50\%$ as
possible, with only small excursions around this value. For given $NPT$
conditions, the relevant thermodynamic averages are computed over a
trajectory of length ranging from $5\times 10^4$ to $10^5$ sweeps. 

In order to locate the melting line, we generate a series of isobaric
paths starting, at very low $T$, from perfect crystals, which are then
heated gradually until melting occurs. This is evidenced by the abrupt
change in e.g. the energy (see Fig.~\ref{fig1}) as well as by the
rounding off of the peaks of the radial distribution function (RDF).
In fact, by this so-called ``heat-until-it-melts'' (HUIM) approach only
the temperature $T^+$ of maximum solid overheating is calculated, which
might be considerably larger than the melting temperature $T_m$~\cite{luo}.
Similarly, the maximum fluid supercooling temperature $T^-$ is
generally far from $T_m$, in fact farther than $T^+$. For a Lennard-Jones
system, the authors of Ref.\,\cite{luo} find that the extent of maximum
overheating/supercooling is only weakly pressure-dependent; moreover,
they suggest the empirical formula $T_m=T^++T^--\sqrt{T^+T^-}$ for
extracting $T_m$ from the boundaries of the hysteresis loop. Another
possibility, which we prefer because it does not rely on a specific
system, is to appeal to the Landau theory of weak first-order
transitions~\cite{chaikin}, which gives the relation
$T_m=(T^-+8T^+)/9$. In all cases here examined, we have verified that
the deviation of $T^+$ from the Landau-type estimate of $T_m$ is small
(6\% at the most) and almost insensitive to pressure.
This indicates that the overall shape of the coexistence curve, which
we are more interested in, is correctly got by the simple HUIM method.

In addition, we calculate the pair excess entropy for each state point:
\be 
s_2=-\frac{k_B}{2}\rho\int{\rm d}{\bf r}\,\left[g(r)\ln
g(r)-g(r)+1\right]\,,
\label{s2}
\ee 
where $k_B$ is Boltzmann's constant, $\rho$ is the number density, and
$g(r)$ is the RDF. $-s_2$ effectively characterizes the
degree of pair translational order in the fluid, as such providing a good
indicator of the melting transition, much better than the density which has
a larger noise-to-signal ratio (see Fig.~\ref{fig1}, bottom
panel)~\cite{saija1}. 

An independent estimate of the location of the melting line is obtained by
the Lindemann criterion~\cite{lindemann,gilvarry}.
The Lindemann ratio $L$ is defined as the mean root square displacement of
the particles about their equilibrium lattice positions, divided by the
nearest-neighbour distance $a$:
\be
L=\frac{1}{a}\left\langle\frac{1}{N}\sum_{i=1}^N\left(
\Delta{\bf R}_i\right)^2\right\rangle^{1/2}\,,
\label{lindratio} 
\ee
\noindent
where the brackets $\left\langle\cdots\right\rangle$ denote an average over
the Monte Carlo trajectory.
The Lindemann rule states that the crystal melts when $L$ becomes larger
than some threshold value $L_c$, which is known to be $0.15-0.16$ for a
fcc solid and $0.18-0.19$ for a bcc solid~\cite{meijer,saija2}. As we shall
see below, the results obtained by the HUIM approach and by the Lindemann
rule compare fairly well with each other.

\section{Results and Discussion}

We first computed the exp-6 phase diagram for $\alpha=10$ (see
Fig.~\ref{fig2}). As anticipated, we identify the melting temperature
$T_m$ with the temperature $T^+$ of maximum solid overheating, assuming
that the difference between the two is indeed minute in relative terms
and almost the same at all pressures.
By comparing the phase diagram of Fig.~\ref{fig2} with that for $\alpha=11$,
reported in Ref.\,\cite{malescio2}, we observe that the overall shape of
the melting line is similar, with the fluid-solid coexistence line passing
through a maximum at temperature $T_M$ and pressure $P_M$.
When increasing pressure at a sufficiently low temperature $T<T_M$,
the initially fluid
system becomes denser and denser until it crystallizes into a fcc solid.
Upon increasing $P$ further, the fcc solid undergoes a transition to a bcc
solid. This transition is related to a decrease in the mean nearest-neighbor
distance with increasing pressure, which brings particles to experience
inner regions of the interaction potential where the repulsion becomes
softer. As the pressure further increases, the bcc solid undergoes
reentrant melting into a denser
fluid. At very high pressures, far beyond the region shown in Fig.~\ref{fig2},
the fluid eventually crystallizes into a fcc hard-sphere-like solid.
In the fluid region above the reentrant melting line, decreasing
temperature at constant pressure leads first to system compression,
and then, contrary to standard behaviour, to an expansion
for further cooling (Fig.~\ref{fig3}). The locus of points where
the density attains its maximum value encloses a region where the
density behaves anomalously (Fig.~\ref{fig2}). Within this region,
open local structures are more favoured than compact ones, causing a
diminution of the number of particles within a given volume with
decreasing $T$. A similar density anomaly has been 
observed in a number of substances (water being the most familiar)
as well as in model
systems characterized by a soft repulsion~\cite{jagla,sadr,kumar,deoliveira3}.

We computed the RDF at a temperature slightly larger than $T_M$, in the
pressure range where reentrant melting occurs (Fig.~\ref{fig4}).
At low pressure, the soft repulsion is quite effective and particles
cannot stay too close to each other. Upon increasing pressure
at constant temperature, more and more particles are able to overcome
the soft repulsion, thus giving origin to a peak close to the hard core
whose height increases
with pressure. Meanwhile the second and third peak become lower,
reflecting the loss of efficacy of the soft-repulsive length scale.
Thus, an increase of pressure causes the gradual turning off of 
the soft-repulsive length scale in favour of the smaller
length scale associated with the inner core. The observed behaviour
differs significantly from that of simple fluids, where all the peaks of
$g(r)$ get higher as pressure grows at constant $T$.

We now follow a different approach for varying the softness of the
exp-6 potential, one in which the parameter $\alpha$ is left unchanged.
An important feature of the exp-6 potential, with regard to its soft nature,
is the existence of a range of interparticle distances where the repulsive
force decreases as two particles get closer. This interval corresponds to
the concave part of the potential, which, in the original exp-6 potential,
extends from $\sigma_M$ up to the inflection point $\sigma_F$.
In the exp-6 model, the value $r_{cross}$ of the interparticle distance
where the repulsion changes from hard-core to soft is $\sigma_M$.
We here instead allow $r_{cross}$ to increase progressively, thus shrinking
the interval where the potential is downward concave, until this interval
disappears when $r_{cross}$ reaches $\sigma_F$ (Fig.~\ref{fig5}).
In this way we are able to vary the relative importance of the hard and soft
components of the exp-6 repulsion.

We have calculated the melting line for a number of ``modified'' exp-6
potentials with $\alpha=10$ by the HUIM approach, see Fig.~\ref{fig6}.
For increasing $r_{cross}$, the portion of melting curve preceding
the maximum
(i.e., for $P<P_M$) remains substantially unaltered, while that having negative
slope becomes more and more flat until, for a value of $r_{cross}$ slightly
smaller than
$\sigma_F$, the slope ${\rm d}T/{\rm d}P$ becomes everywhere positive and the
melting-curve maximum disappears. Upon further increasing $r_{cross}$,
the part of melting line for $P>P_M$ becomes steeper and steeper. This
behaviour clearly underlines the fundamental role played by the soft-repulsive
component of the potential in giving origin to
reentrant melting. We stress that the reentrant region in fact disappears
for a value of $r_{cross}$ being a little smaller than $\sigma_F$, that is
when the potential has still an interval of downward concavity. This suggests
that the existence of a concave repulsive region in the potential,
while being crucial for reentrant melting, is not strictly
sufficient for its occurrence.
This result is consistent with the findings of a recent investigation
of the phase behaviour of a potential consisting in a smoothened
hard-core plus a repulsive step, showing that a melting line with
a maximum and a reentrant portion occurs only for a sufficiently
wide repulsive shoulder~\cite{fomin}.

As $r_{cross}$ increases, the region of anomalous density behaviour becomes
less evident until it disappears completely, at least in the stable
fluid phase, when the slope $dT/dP$ becomes everywhere positive
(Fig.~\ref{fig7}). Actually, with increasing $r_{cross}$, the density
anomaly migrates to lower temperatures, while the melting line moves
to high temperatures. As a result, the line of maximum density is
swallowed up by the solid phase (in fact, for sufficiently large $r_{cross}$,
the density maximum disappears completely). A similar behaviour has recently
been observed for a family of isotropic pair potentials with
two repulsive length scales~\cite{deoliveira4}. The close relation between the
occurrence of reentrant melting and that of the density anomaly (be it in the
stable or metastable fluid phase) points to a common origin for the two.
Indeed, it appears that a prerequisite for both phenomena
is the existence of two repulsive length scales, which in turn gives
rise to two different patterns of local order in the system. Therefore,
the crossover between low-density/low-temperature open structures and
high-density/high-temperature compact ones is at the basis of the
remelting of the solid into a denser fluid as well as of
the decrease of the density upon isobaric cooling.

\section{Conclusions}

In this paper, we have studied how the melting behaviour of a system of
particles interacting through the exp-6 potential depends on the
repulsion softness. We find that, by varying the softness parameter
$\alpha$, the anomalous features of the phase diagram do not change
qualitatively while the typical pressure and temperature where reentrant
melting occurs change considerably. As the repulsion gets softer,
i.e., as $\alpha$ decreases, the reentrant-melting region moves to smaller
pressures and temperatures.
In fact, a softer repulsion and the associated length scale lose
efficacy more rapidly with increasing pressure and temperature.

We have shown that the feature of the exp-6 potential 
that is crucial for the occurrence of anomalous melting is the existence
of a concave region in the repulsive part of the potential.
By progressively reducing the extent of this interval, in fact, the 
negatively-sloped portion of the melting line as well as the $P$-$T$
region where the density is anomalous tend to disappear. This is an 
indication of the close relation between reentrant melting and 
density anomaly, both phenomena being linked to the turning on and off
of the two repulsive length scales in the system.

In spite of the modellistic nature of the system investigated,
the sensitive dependence of the anomalous-melting region on the
steepness of the repulsive interaction has a counterpart in real systems.
By looking at the phase behaviour of elements displaying anomalous melting,
it is possible to observe that the pressures and temperatures where 
the anomalies occur greatly vary from one element to the other.
For example, the maximum in the melting line is at about 2 GPa for
Cs~\cite{jayaraman}, 30 GPa for Na~\cite{gregoryanz}, and is predicted
around 100 GPa for H~\cite{bonev}. These experimental and theoretical
results can be rationalized by considering that atoms with more electrons
are more susceptible, at least within the same chemical family, to
pressure-induced structural softening. Besides alkali metals, this
trend is also observed for rare gases~\cite{boehler}.

\clearpage

\begin{figure}
\caption{
Modified exp-6 potential with $r_{cross} = 0.5$ at $P=1000$:
The bcc-fluid hysteresis loop for the excess energy (top panel)
and the two-body entropy~\cite{saija1} (bottom panel) at $P=1000$.}
\label{fig1}
\end{figure}

\begin{figure}
\caption{(Color online). Phase diagram of the exp-6 model for $\alpha=10$.
Pressure $P$ and temperature $T$ are expressed in units of $\epsilon/\sigma^3$
and $\epsilon/k_B$, respectively.
Melting points, located by the Lindemann criterion, are represented as full
green dots. The boundary between the bcc and fcc solids (black dotted line)
is roughly obtained by drawing a straight line from the full square at $T=0$
(obtained through an exact total-energy calculation) to the square, lying
on the melting curve, where the value of the Lindemann ratio switches
from $0.15-0.16$ (fcc) to $0.18-0.19$ (bcc)~\cite{meijer,saija2}.
We also plot the fluid-solid coexistence locus as obtained by the
HUIM criterion (blue open dots), which indeed agrees well
with the Lindemann-based estimate.
The locus of density maxima in the fluid phase is marked by red diamonds.
All lines in the figure are guides to the eye.}
\label{fig2}
\end{figure}

\begin{figure}
\caption{Exp-6 model for $\alpha=10$: Fluid number density $\rho$ (in units
of $\sigma^{-3}$) as a function of temperature for $P =800,900,1000,1200$
(full dots). All lines are fourth-order polynomial fits of the data points.}
\label{fig3}
\end{figure}

\begin{figure}
\caption{(Color online). Exp-6 model for $\alpha=10$:
Radial distribution function $g(r)$ for $T=5$ and three pressures,
$P=150$ (red solid line), 750 (blue dashed line), 1500 (green dash-dotted
line).}
\label{fig4}
\end{figure}

\begin{figure}
\caption{(Color online). Modified exp-6 potential for $\alpha=10$.
The vertical lines correspond to different values of $r_{cross}$ (see
text for explanation): $0.4659\approx\sigma_M$ (black solid line), 0.5
(red dashed line), 0.525 (blue dash-dotted line), $0.53442\approx\sigma_F$
(green dotted line), 0.55 (orange double-dash dotted line).
The full green dot marks the point, $\sigma_F$, where the second
derivative of the potential vanishes.}
\label{fig5}
\end{figure}

\begin{figure}
\caption{(Color online). Phase diagram of the modified exp-6 potential
for $\alpha=10$ and for several values of $r_{cross}$: $0.4659\approx
\sigma_M$ (black solid line), 0.5 (red dashed line), 0.525
(blue dash-dotted line), $0.53442\approx\sigma_F$ (green dotted line),
0.55 (orange double-dash dotted line). The lines being shown are polynomial
fits of the simulation data (full symbols).}
\label{fig6}
\end{figure}

\begin{figure}
\caption{Modified exp-6 potential for $\alpha=10$ and $r_{cross}=0.5$:
Reduced number density of the fluid as a function of temperature for
$P=1000$ and 1200 (full triangles). All lines are fourth-order polynomial fits of the
data points.}
\label{fig7}
\end{figure}

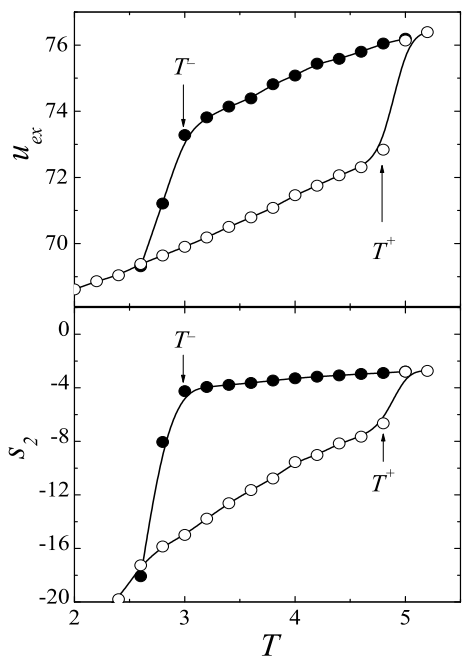
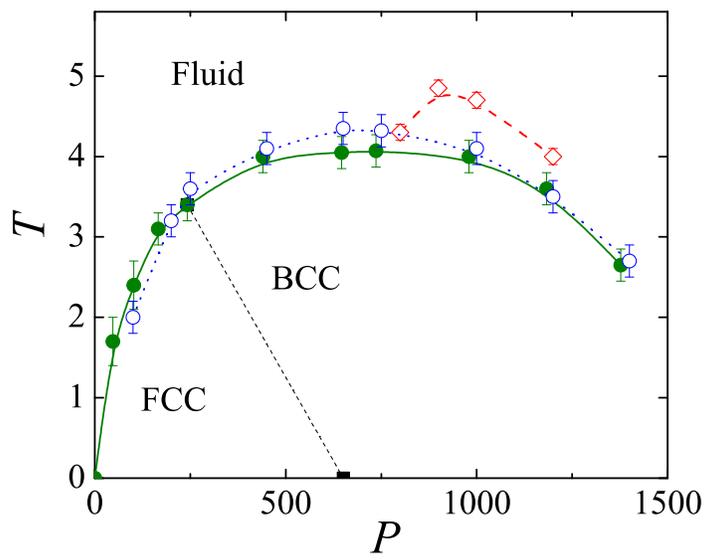
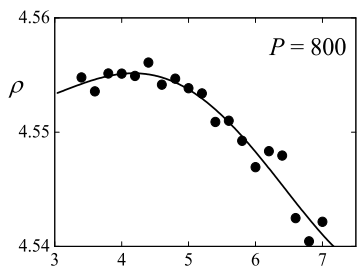
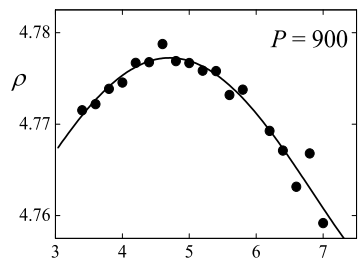
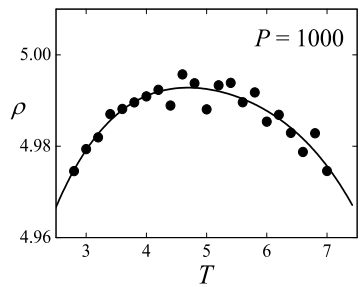
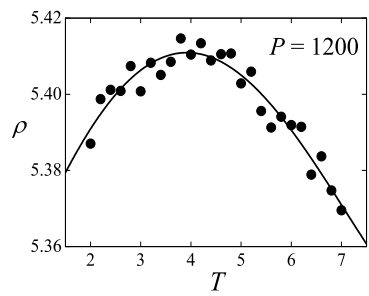
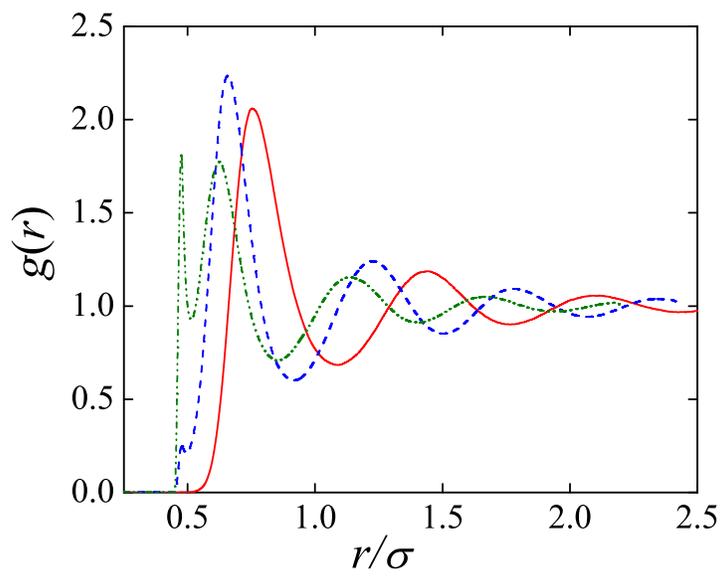
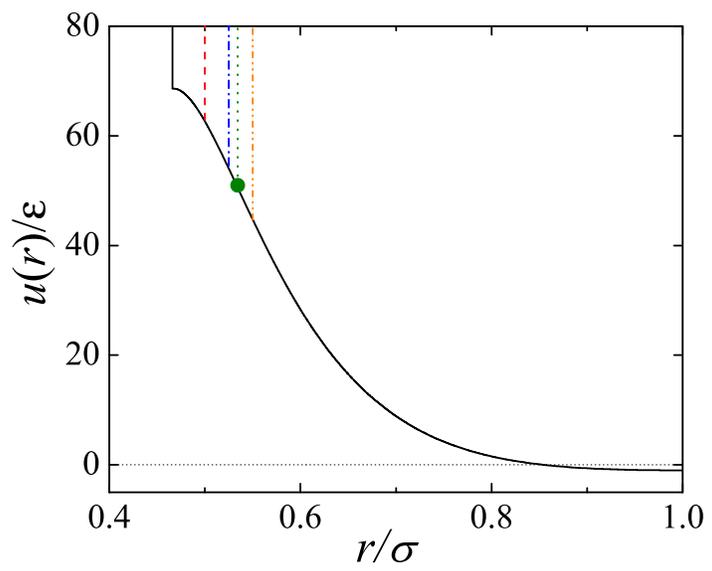
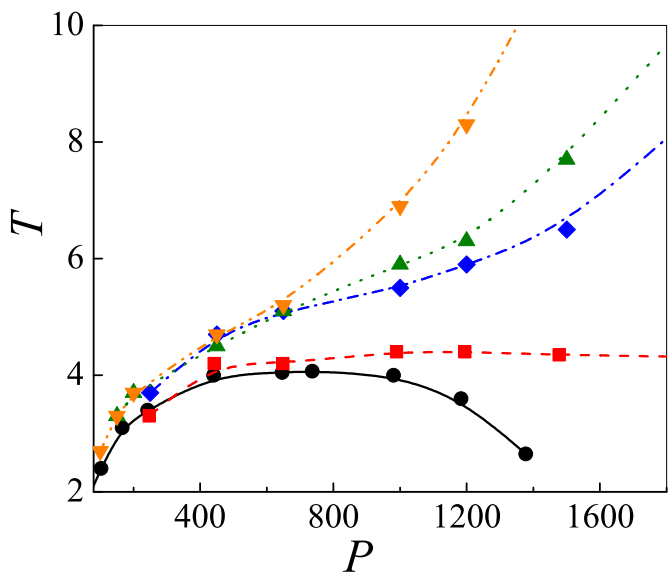
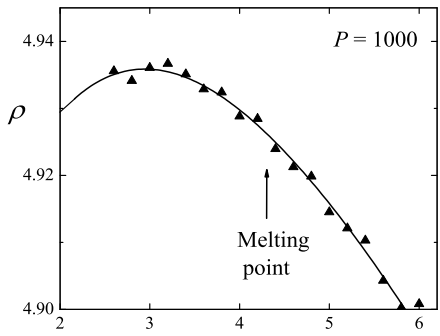
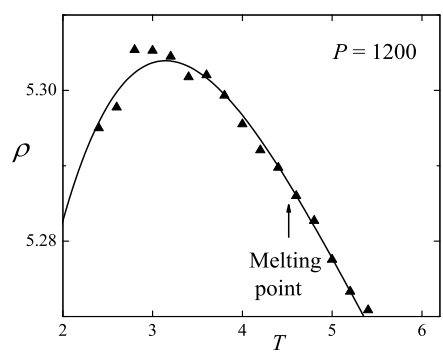
\end{document}